\documentclass[12pt]{article}
\usepackage{amssymb}
\usepackage{enumerate}
\usepackage{dcolumn}

\usepackage{graphicx}
\usepackage{dcolumn}
\usepackage{bm}

\begin{document}


\def\H{{\cal H}}
\def\ttheta{\tilde{\theta}}

\def\beq{\begin{equation}}
\def\eeq{\end{equation}}
\def\bea{\begin{eqnarray}}
\def\eea{\end{eqnarray}}
\def\ben{\begin{enumerate}}
\def\een{\end{enumerate}}
\def\la{\langle}
\def\ra{\rangle}
\def\a{\alpha}
\def\b{\beta}
\def\g{\gamma}\def\G{\Gamma}
\def\d{\delta}
\def\e{\epsilon}
\def\phi{\varphi}
\def\k{\kappa}
\def\l{\lambda}
\def\m{\mu}
\def\n{\nu}
\def\o{\omega}
\def\p{\pi}
\def\r{\rho}
\def\s{\sigma}
\def\t{\tau}
\def\L{{\cal L}}
\def\S{\Sigma }
\def\gsim{\; \raisebox{-.8ex}{$\stackrel{\textstyle >}{\sim}$}\;}
\def\lsim{\; \raisebox{-.8ex}{$\stackrel{\textstyle <}{\sim}$}\;}
\def\gtrsim{\gsim}
\def\lessim{\lsim}
\def\loc{{\rm local}}
\def\vm{v_{\rm max}}
\def\bh{\bar{h}}
\def\del{\partial}
\def\nab{\nabla}
\def\half{{\textstyle{\frac{1}{2}}}}
\def\fourth{{\textstyle{\frac{1}{4}}}}

\begin{center} {\Large \bf
When is $g_{tt}\, g_{rr} =-1$?}
\end{center}

\vskip 5mm
\begin{center} \large
{{Ted Jacobson\footnote{E-mail: jacobson@umd.edu}}}\end{center}

\vskip  0.5 cm
{\centerline{\it Department of Physics}}
{\centerline{\it University of Maryland}}
{\centerline{\it College Park, MD 20742-4111, USA}}

\vskip 1cm

\begin{abstract}
The Schwarzschild metric, its Reissner-Nordstrom-de Sitter
generalizations to higher dimensions, and some further
generalizations all share the feature that $g_{tt}\, g_{rr}=-1$
in Schwarzschild-like coordinates.  In this pedagogical note
we trace this feature to the condition that
the Ricci tensor (and stress-energy tensor in a
solution to Einstein's equation)
has vanishing radial null-null component,
i.e.\  is proportional to the metric
in the $t$-$r$ subspace.
We also show this condition holds if and only if the
area-radius coordinate is an affine parameter on
the radial null geodesics.
\end{abstract}


A notable feature of the Schwarzschild solution to
Einstein's equation, which
generalizes to a rather wide class of static
solutions of the form
\beq
ds^2 = -f(r) \, dt^2 + f(r)^{-1}\, dr^2 + r^2\, h_{ij}(x) dx^i dx^j,
\label{form1}
\eeq
 is the fact that the metric
component $g_{rr}$ is the reciprocal of $-g_{tt}$. The
purpose of this pedagogical note is to point out that
this commonly occuring feature arises if and only if
the radial null-null components of the Ricci tensor
(which are equal) vanish; equivalently, if the
restriction of  the Ricci tensor to the $t$-$r$
subspace is proportional to $g_{\mu\nu}$. When the
Einstein equation is satisfied this condition holds
for the stress-energy tensor,
implying that the radial
pressure is the negative of the energy density.
 An equivalent condition
is that the coordinate $r$ is an affine parameter on
the radial null geodesics. Oddly, the following
elementary observations are not easy to find in the
literature,
although they have surely been noticed many times
before.
An example is Ref.~\cite{Guendelman:1996pg}, in which
it is shown that the metric form (\ref{form1}) follows
from this condition on the stress-energy tensor.

Before explaining the feature, let me cast it
in the broadest context for static solutions to
Einstein's equation.
It applies in spherical symmetry
with either Maxwell electrodynamics
(the Reissner-Nordstrom solution) or
Born-Infeld nonlinear
electrodynamics~\cite{Demianski:1986wx},
and persists in the presence of a cosmological constant.
In the Maxwellian case, one has
$f(r)=1-2M/r + Q/r^2 - \Lambda r^2/3$,
and the transverse metric $h_{ij}$ describes a unit
2-sphere.
The 2-sphere can be replaced by a
flat metric or a metric of constant negative curvature,
simultaneously replacing the initial 1 in
$f(r)$ by $k=$ 0 or $-1$ respectively. This example
generalizes to $d$ spacetime dimensions as well,
in which case  $h_{ij}$ is any metric satisfying
$R_{ij}(h) = (d-3)k h_{ij}$~\cite{Birmingham:1998nr},
which need not have constant curvature.
The special form (\ref{form1}) also applies
to a spherically symmetric global monopole
(``hedgehog") configuration of an O(3) nonlinear
sigma model~\cite{Barriola:1989hx}, which
is equivalent~\cite{Guendelman:1991qb}
to a ``string hedgehog," i.e.\ a spherically symmetric
continuous distribution of radial strings.

Now let us consider a static metric of the form
\beq
ds^2 = -f(r) \, dt^2 + g(r)\, dr^2 + r^2\, h_{ij}(x) dx^i dx^j,
\eeq
in any number of spacetime dimensions $d$.
The radial null vectors $l^\mu$ can be scaled to
have components
$l^t=g^{1/2}$, $l^r=\pm f^{1/2}$, and $l^i=0$.
The off-diagonal component $R_{tr}$ of the
Ricci tensor vanishes by time reflection symmetry.
Hence the two radial null-null components of the Ricci
tensor are equal, and
given by $R_{\mu\nu}l^\mu l^\nu = g R_{tt}
+ f R_{rr} = (d-2)(fg)'/2rg$, which vanishes if and only if
the product $fg$ is constant. A re-scaling of the time
coordinate brings this to the form $fg=1$.

The 4-velocity of a radial null curve $(t(\l),r(\l),0,0)$
 is $k^\mu=(\dot{t}, \dot{r},0,0)$, with $f \dot{t}^2 =g\dot{r}^2$,
 where the dot stands for $d/d\l$.
 Spherical symmetry implies this curve is a geodesic,
and if $\l$ is an affine parameter then the
$t$ translation symmetry implies that
the ``energy"  $g_{t\mu} k^\mu=f \dot{t}=\sqrt{fg}\, \dot{r}$
is constant. Thus $fg$ is constant
if and only if $r$ is linearly related to $\l$, and hence
is also an affine parameter.

The equivalence of the two conditions
for constancy of $fg$ can be seen directly
using the Raychaudhuri equation
for the expansion of the radial null geodesic
congruence,
\beq
\dot{\theta}=-\frac{1}{d-2}\theta^2 - R_{\mu\nu}k^\mu k^\nu.
\label{Ray}
\eeq
Here the expansion
$\theta=d(\ln \d A)/d\l$ is the fractional rate of
change of a transverse area element $\d A$ with
respect to the affine parameter, and
$k^\mu$ is the affinely parametrized velocity vector.
The twist and shear terms generally present
in the Raychaudhuri equation have been omitted since
they vanish for this congruence.
Since the area element scales simply as $r^{d-2}$
along the congruence, we have $\theta=(d-2)\dot{r}/r$,
hence $\dot{\theta}=-(d-2)\dot{r}/r^2 + (d-2)\ddot{r}/r$.
Inserting these into (\ref{Ray}) yields
\beq
\ddot{r}=-\frac{r}{d-2}R_{\mu\nu}k^\mu k^\nu.
\label{Ray2}
\eeq
Hence $R_{\mu\nu}k^\mu k^\nu=0$ if and only if
$r$ is an affine parameter on the radial null geodesics.

To close let us return to the condition on the stress tensor.
When the Einstein equation holds,
$R_{\mu\nu}k^\mu k^\nu=0$ is equivalent to
$T_{\mu\nu}k^\mu k^\nu=0$.
This condition on the stress tensor
of course holds in vacuum, and with
a cosmological constant since then
$T_{\mu\nu}\propto g_{\mu\nu}$
and $g_{\mu\nu}k^\mu k^\nu=0$.
As to electromagnetic fields, the Maxwell
and Born-Infeld stress tensors
are both a sum of terms proportional to
$F_{\mu\sigma}F_\nu{}^\sigma$ and the metric
$g_{\mu\nu}$.
When contracted with $k^\mu k^\nu$
the metric term contributes nothing, while the first
term yields $V_\sigma V^\sigma$, where
$V_\sigma = k^\mu F_{\mu\sigma}$.
For a radial magnetic field, $F_{\mu\nu}$ is tangential,
so $V_\sigma=0$.  For a radial electric field,
$F_{\mu\nu} \propto k_{[\mu}l_{\nu]}$, where
$l_\nu$ is the other radial null direction. Hence
$V_{\sigma}\propto k_\sigma$,
so $V_\sigma V^\sigma=0$.
For the nonlinear O(3) sigma model with an
isovector field $\phi^i$ of fixed norm,
the stress tensor is proportional
to $\phi^i_{,\mu}\phi^i_{,\nu} - g_{\mu\nu}
(\half g^{\rho\sigma}\phi^i_{,\rho}\phi^i_{,\sigma})$.
For the hedgehog configuration
the partial derivatives with respect to $t$ and $r$
vanish, hence the $t$-$r$ part of the
stress tensor is proportional
to the metric and therefore satisfies the condition.
In four spacetime dimensions, the transverse
stress vanishes, so this is equivalent to a
string hedgehog. In other dimensions, the
string hedgehog is a distinct solution of this
type, with vanishing transverse stress.
For most other
forms of matter,  $T_{\mu\nu}k^\mu k^\nu$ does not vanish.
For example, it does not vanish
in the fluid interior of a stellar solution, in
scalar-tensor solutions, or in non-Abelian Einstein-Yang-Mills
black hole spacetimes.

\section*{Acknowledgments} I am grateful to
E. Guendelman for alerting me to some relevant
references. This work was supported in part by the
National Science Foundation under grant PHY-0601800.

\end{document}